\documentclass[fleqn,usenatbib]{mnras}
\usepackage{newtxtext,newtxmath}
\usepackage[T1]{fontenc}

\DeclareRobustCommand{\VAN}[3]{#2}
\let\VANthebibliography\thebibliography
\def\thebibliography{\DeclareRobustCommand{\VAN}[3]{##3}\VANthebibliography}

\usepackage{graphicx}	% Including figure files
\usepackage{amsmath}	% Advanced maths commands
\usepackage{amssymb}	% Extra maths symbols

\hypersetup{linkcolor=red,citecolor=blue,filecolor=cyan,urlcolor=magenta}

\newcommand{\gmax}{\gamma_{\rm max}}
\newcommand{\gf}{\gamma_{9}}
\newcommand{\beq}{\begin{equation}}
\newcommand{\eeq}{\end{equation}}
\newcommand{\drj}{\Delta r_{\rm j}}
\newcommand{\drjc}{\Delta r_{\rm j,2}}
\newcommand{\rjc}{ r_{\rm j,2}}
\newcommand{\bc}{B_{-5}}

\newcommand{\td}{\tau_{\rm dyn}}
\newcommand{\tsc}{\tau_{\rm sc}}
\newcommand{\tsh}{\tau_{\rm sh}}

\newcommand{\tes}{\tau_{\rm esc}}

\title[Shear acceleration in large-scale jets]{Particle acceleration in shearing flows: the case for large-scale jets}
\author[Wang et al.]{
Jie-Shuang Wang,$^{1}$\thanks{jiesh.wang@gmail.com}
Brian Reville,$^{2}$
Ruo-Yu Liu,$^{3,4}$
Frank M. Rieger,$^{2,5}$
Felix A. Aharonian$^{2,6,7}$\\
$^{1}$Tsung-Dao Lee Institute, Shanghai Jiao Tong University, Shanghai 200240, China\\
$^{2}$Max-Planck-Institut f\"ur Kernphysik, Saupfercheckweg 1, D-69117 Heidelberg, Germany\\
$^{3}$School of Astronomy and Space Science, Nanjing University, Nanjing 210093, China\\
$^{4}$Key laboratory of Modern Astronomy and Astrophysics (Nanjing University), Ministry of Education, Nanjing 210023, China\\
$^{5}$Institute of Theoretical Astrophysics, University of Heidelberg, Philosophenweg 12, D-69120 Heidelberg, Germany\\
$^{6}$Dublin Institute for Advanced Studies, 31 Fitzwilliam Place, Dublin 2, Ireland\\
$^{7}$High Energy Astrophysics Laboratory, RAU, 123 Hovsep Emin St Yerevan 0051, Armenia
}

\date{}
\pubyear{}

\begin{document}

\label{firstpage}
\pagerange{\pageref{firstpage}--\pageref{lastpage}}
\maketitle

\begin{abstract}
X-ray observations of kilo-parsec scale jets indicate that a synchrotron origin of the sustained non-thermal emission is likely. This requires distributed acceleration of electrons up to near PeV energies along the jet. The underlying acceleration mechanism is still unclear. Shear acceleration is a promising candidate, as velocity-shear stratification is a natural consequence of the collimated flow of a jet. We study the details of shear acceleration by solving the steady-state Fokker-Planck-type equation and provide a simple general solution for trans-relativistic jets for a range of magnetohydrodynamic turbulent power-law spectra. In general, the accelerated particle population is a power-law spectrum with an exponential-like cut-off, where the power-law index is determined by the turbulence spectrum and the balance of escape and acceleration of particles.
Adopting a simple linearly decreasing velocity profile in the boundary of large-scale jets, we find that the multi-wavelength spectral energy distribution of X-ray jets, such as Centaurus A and 3C 273, can be reproduced with electrons that are accelerated up to $\sim$ PeV. In kpc-scale jets, protons may be accelerated up to $\sim$ EeV, supporting the hypothesis that large-scale jets are strong candidates for ultra-high-energy-cosmic-ray sources within the framework of shear acceleration. 
\end{abstract}


\section{Introduction}\label{section:intro}

Jets of active galactic nuclei (AGNs) can transport mass and energy from accreting 
super-massive black holes at the galaxy's center, feeding large-scale lobes that shine 
across the electromagnetic spectrum. In Fanaroff–Riley (FR) type galaxies, giant radio 
lobes, inflated by jets, are observed even on Mpc-scale \citep{Fanaroff1974}, implying 
that they can influence the intracluster medium significantly. 
In FR I jets, the region with higher radio-brightness is closer to the nucleus, while 
in FR II sources, the jet typically terminates at a working surface far from the core. 
The large-scale FR jets can also be bright in optical and X-ray bands, which makes them perfect candidates to explore jet dynamics, energy dissipation, and radiation processes 
\citep[see][for a recent review]{Blandford2019}. 
To date, approximately one hundred jets have been observed in X-rays \citep[see][for reviews]{Harris2006,Schwartz2015}. 
Although the dominant structures are bright knots and terminal hotspots, quasi-continuous emission is observed in some X-ray jets, such as Centaurus A \citep{Kraft2002,Kataoka2006}. 

In FR I jets, the radio, optical, and X-ray spectrum can typically be explained by synchrotron radiation from a single population of electrons \citep[e.g.,][]{Perlman2001,Hardcastle2001,Sun2018}. 
The X-ray measurements of FR II jets can however exhibit much harder spectra \citep[e.g.,][]{Jester2006,Jester2007}. 
This has motivated studies suggesting that in FR II jets X-rays are produced via inverse Compton (IC) scattering by low-energy electrons (around tens of MeV) off cosmic microwave background (CMB) photons, while relativistic jets are required with bulk Lorentz factors of order $10$ \citep{Tavecchio2000,Celotti2001}.
Although both emission mechanisms can account for the X-ray spectral energy distribution (SED), they have different predictions.
Comparing with a synchrotron radiation model, the IC/CMB model will lead to a higher gamma-ray flux, but less polarization \citep[e.g.][]{Uchiyama2008}. Corresponding tests have been performed and favour the synchrotron origin \citep[see][for recent reviews]{Georganopoulos2016,Perlman2020}, which includes the optical polarimetry observation of PKS 1136-135 \citep{Cara2013}, 3C 273, PKS 0637-752, and PKS 1150+497 \citep{Perlman2020}, and gamma-ray studies of 3C 273 \citep{Meyer2014}, PKS 0637-752 \citep{Meyer2015}, PKS 1136–135, PKS 1229–021, PKS 1354+195, and PKS 2209+080 \citep{Breiding2017}.
With the IC origin of X-ray emission in these sources now disfavoured, it is reasonable to explore a synchrotron origin in other jets and its physical implications.
The electrons radiating in X-rays suffer substantial synchrotron cooling losses on a timescale of 
decades, such that for a localised acceleration site, the X-ray-bright region should not extend more than tens of parsecs, contrary to observations. 
As such, a distributed {\em in-situ} (re-)acceleration mechanism is required for large-scale X-ray jets. 

Several candidate acceleration mechanisms have been proposed for particle acceleration in jets 
\citep[see][for recent reviews]{Blandford2019,Matthewsetal}. 
Shock acceleration, being arguably the best understood, is commonly invoked as the primary 
acceleration mechanism in jets, motivated by its successful application in many other astrophysical scenarios, such as 
supernova remnants \citep[e.g.,][]{Drury1983,Blandford1987,Bell2013}. 
While acceleration of electrons to high energies at shocks is possible, it only occurs where strong shock formation is possible, for example re-collimation shocks or the jet termination shock. For magnetically dominated jets, relativistic magnetic reconnection, possibly driven by the
kink instability \citep[e.g.][]{Begelman1998}, has been suggested as a driver for particle acceleration. This process has 
been demonstrated to be efficient in particle-in-cell (PIC) simulations
\citep[e.g.,][]{Sironi2014,Werner2016}. 
However, beyond the kpc-scale, jets are generally kinetically dominated \citep[e.g.,][]{Sikora2005,Potter2017,Chatterjee2019}, 
and as such, an alternative mechanism that can accelerate particles continuously over tens of kiloparsecs is needed.
Stochastic acceleration in a turbulent jet boundary layer is a possible alternative \citep[e.g.,][]{Stawarz2002}, though rapid acceleration requires a high Alfv{\'e}n speed in the jet \citep{OSullivan}. For typical conditions of $1-100\,$kpc scale jets, \citet{Liu2017} argue that stochastic acceleration may not be efficient enough to overcome the cooling and escape of particles, and hence may have difficulties to produce X-ray emitting electrons.

Shear acceleration, another type of Fermi acceleration mechanism, can in principle accelerate high-energy electrons in the steep velocity gradients that inevitably develop at the boundary of jets (see e.g., \citealt{Berezhko1981, BerezhkoKrymskii1981, Earl1988} for early studies and \citealt{Rieger2019Galax} for a recent review). 
Indeed, high-resolution radio imaging and polarization studies of large-scale jets indicate the presence of such velocity gradients transverse to the main jet axes \citep[e.g.,][]{Laing2014,Gabuzda2014,Nagai2014,Boccardi2016}. 
Numerical simulations have further shown that the global stability of jets also depends on the properties of the shear layer \citep[e.g.,][]{Mizuno2013,Kim2018}. 
In gradual shearing flows, particles gain/lose energy by elastically scattering off magnetic field inhomogeneities \citep[e.g.,][]{ Rieger2016,Webb2018}, which are considered to be frozen-in to the corresponding velocity-shear layers, i.e. the particle energy is approximately conserved in the local fluid frame.
For shear acceleration in kpc-scale jets, under favourable conditions, the accelerated particles are expected to form power-law spectra, with cut-offs predicted at $\sim$ EeV for protons, although cooling limits electrons to $\sim$ PeV energies \citep[e.g.,][]{Rieger2019ApJ,Rieger2021}. To this end, \citet{Liu2017} have shown that the accelerated electrons can in principle emit X-ray up to 100\,keV with a hard spectrum via synchrotron radiation, while the diffusive escape of accelerated electrons can soften the spectrum. The joint effect of shear acceleration and diffusive escape results in a large range for the accelerated electron spectral slope, but the influence on the predicted X-ray spectrum in large-scale jets has yet to be quantified. 

In this paper, we explore particle acceleration in the gradual shear flows at the boundary of large-scale jets by solving a Fokker–Planck-type equation with an analytical approach, where the acceleration, cooling, and diffusive escape terms are included simultaneously.
The steady-state solution is derived in Section \ref{section:Shear_acc}.
In Section \ref{section:X-ray-jet}, we study the corresponding radiation from the accelerated particles and apply it to the observed large-scale jets. Exemplary modellings of the SEDs in Centaurus A and 3C 273 are presented.
In Section \ref{section:CR}, we explore the capability of acceleration of cosmic rays (CRs) in large-scale jets. The conclusion and discussion are presented in Section \ref{section:conclusion}.

\section{Accelerated particle spectra in shearing flows }\label{section:Shear_acc}

In fast shearing flows, particles can gain energy by elastically scattering off magnetic field
inhomogeneities embedded in the local fluid. This can be understood as a stochastic 
Fermi-type acceleration mechanism \cite[e.g.][]{Rieger2007}. 
We consider first electrons. Taking these to be magnetised, the mean scattering time can be formulated as \citep[see for example][]{SchlickeiserBook},
\begin{equation}
\tau_{\rm sc} = \xi_1^{-1} \left(\frac{r_{\rm L}}{\Lambda_{\rm max}}\right)^{1-q}
\frac{r_{\rm L}}{c}\equiv A_0\gamma^{2-q},\label{tscatter}
\end{equation}
where $\xi_1=\delta B^2/B_0^2$ denotes the energy density ratio of turbulent field ($\delta B$) 
to the mean magnetic field ($B_0$), $r_{\rm L} = \gamma m_{\rm e} c^2/eB_0$ is the Larmor radius of an electron with Lorentz 
factor $\gamma$ and $A_0= \xi_1^{-1}(\Lambda_{\rm 
max}/c)^{q-1}(m_ec/eB_0)^{2-q}$. The index $q$ denotes the power-law index of the turbulent spectrum, 
e.g., $q = 5/3$ for a Kolmogorov-type turbulence, $q = 3/2$ for Kraichnan-type, and $q=1$ for Bohm-type.
$\Lambda_{\rm max}$ is the outer turbulence scale. If the turbulent field dominates over the mean field, provided $q>1$, it is reasonable to replace the mean field with the strength at the largest scales, i.e.  take $B_0 \approx \delta B$ and $\xi_1 \approx 1$. We here focus on the range of 
$1\leq q\leq2$, and note that Eq. (\ref{tscatter}) only applies if $r_{\rm L}<\Lambda_{\rm max}$. 
Beyond this scale, the scattering time will follow a different behaviour. 

The energy space diffusion coefficient in a gradual shear flow is given by \citep[e.g.,][]{Rieger2006}
\begin{equation}
\left \langle \frac{\Delta \gamma^2}{\Delta t} \right\rangle_{\rm sh} \simeq  
D_{\rm sh}\gamma^2\tau_{\rm sc}\equiv A_1\gamma^{4-q},
\end{equation}
where $\left<...\right>$ denotes averaging over an isotropic particle distribution.
For a cylindrical outflow, the viscous momentum transfer coefficient is \citep[e.g.,][]{Rieger2004, Webb2018},
\beq
D_{\rm sh}=\frac{1}{15} \Gamma_{\rm j}^4(r) c^2 \left(\frac{\partial 
\beta_{\rm j}(r)}{\partial r}\right)^2\,,
\label{Diffu_shear}
\eeq
where $\Gamma_{\rm j}(r)=(1-\beta_{\rm j}^2)^{-1/2}$ and $\beta_{\rm j}(r)c$ are the bulk 
Lorentz factor and velocity of different layers located at a distance $r$ from the jet axis, 
and $A_1=A_0D_{\rm sh}$. 
For a trans-relativistic flow and a linearly decreasing velocity profile, this expression essentially becomes independent of $r$.

The corresponding acceleration time is,
\begin{equation}
\tau_{\rm sh} = {2\over 6-q}D_{\rm sh}^{-1}\tau_{\rm sc}^{-1}={2\over 6-q}A_1^{-1}\gamma^{q-2}. \label{tacc}
\end{equation}

In magnetised jets, the high-energy electrons will undergo continuous energy losses via 
synchrotron and IC radiation. 
Neglecting Klein-Nishina corrections, the cooling rate is 
\beq
\langle \dot{\gamma}_{\rm c}\rangle = {\sigma_T\gamma^2B^2\over6\pi m_ec}
%{4\gamma^2e^4B^2\over9m_e^3c^5}
(1+f)\equiv A_2 \gamma^2,\label{tcool}
\eeq
where $\sigma_T$ is the Thomson scattering cross-section, and $f = U_{\rm rad}/U_B$ is the energy 
density ratio between the target photon field ($U_{\rm rad}$) for the IC process and the magnetic field ($U_B=B^2/8\pi$). 
For IC/CMB scattering, $U_{\rm CMB}=4.18\times10^{-13}(1+z)^4$ erg cm$^{-3}$ and hence for $B=10^{-5}\bc$ G, we find $f=0.1(1+z)^4\bc^{-2}$. Synchrotron radiation will dominate over IC/CMB if $B>3.2(1+z)^2~\mu$G.
Combining radiative losses and shear acceleration, the maximum energy is obtained from 
$\langle \dot{\gamma}_{\rm c}\rangle=\gamma/\tau_{\rm sh}$ for electrons, which yields
\begin{equation}
\gmax=\left({6-q\over2}{A_1\over A_2}\right)^{1/(q-1)}%=\left[ \frac{9(6-q)}{8\xi_1}{m_e\over e^2}D_{\rm sh}  ({m_ec^2\over eB})^{4-q}\Lambda_{\rm max}^{q-1}  \right]^{1/(q-1)}
.\label{eq:gmaxsyn}
\end{equation}

An energetic particle can also escape from the acceleration zone by diffusion, and the corresponding escape time is given by
\begin{equation}
\tau_{\rm esc}={\drj^2\over2\kappa}=\frac{3\drj^2}{2c^2}\tau_{\rm sc}^{-1}\equiv A_3 
\gamma^{q-2},\label{tesc}
\end{equation}
where $\drj$ is the width of the shearing region of the jet, 
$\kappa=c^2\tau_{\rm sc}/3$ is the particles' spatial diffusion coefficient, and $A_3=1.5\drj^2A_0^{-1}c^{-2}$. Note, that we assume isotropic diffusion. A strong guide field will inhibit escape, but also reduces the acceleration rate \citep{Webb1989}.

Taking all these effects into consideration, the evolution of the particle distribution function $n(\gamma,t)$ can be expressed as a Fokker–Planck-type equation  \cite[e.g.][]{Liu2017}, 
\begin{align}
\frac{\partial n(\gamma,t)}{\partial t}
&=\frac{1}{2}\frac{\partial}{\partial \gamma}\left[\left \langle \frac{\Delta \gamma^2}{\Delta t}\right\rangle\frac{\partial n(\gamma,t)}{\partial \gamma} \right] \label{eq:Fokker} \\
&- \frac{\partial}{\partial \gamma}\left[ \left( \frac{1}{ \gamma}\left\langle \frac{\Delta \gamma^2}{\Delta t}\right\rangle - \langle \dot{\gamma}_{\rm c}\rangle \right) n(\gamma,t)\right]-\frac{n(\gamma,t)}{\tau_{\rm esc}}+Q(\gamma,t)\,, \nonumber  
\end{align}
where $Q(\gamma,t)$ denotes the injection rate. 

From Eq. (\ref{tacc}), it follows that for $q<2$ the acceleration time is a decreasing function 
of particle energy. This means that shear acceleration is in theory more efficient to accelerate 
higher-energy particles, although it likely requires seed particles to be accelerated to sufficiently 
high energies \citep{Stawarz2002,Rieger2006}. 
Several processes have been suggested for the acceleration of seed particles, including stochastic 
(classical 2nd order Fermi) acceleration \citep{Liu2017}, magnetic reconnection resulting from the
Kelvin-Helmholtz instability in the shearing region \citep{Sironi2020}, and shock acceleration \citep{Tavecchio2020}. 
In the latter case, these seed particles remain essentially uncooled for $\gtrsim7.8\times10^4 
\bc^{-2} (1+f)^{-1}$ yrs, and are responsible for the radio-to-near-infrared 
emissions from the jet. 
In this paper, we do not consider a specific acceleration mechanism for the seed particles, 
which might be one or more of the above mechanisms, as we are only interested in particles 
with higher energies accelerated through shear acceleration. Instead, we assume that seed 
particles are injected below a certain energy ($\gamma_{\rm cr}$), so that we can take
$Q(\gamma,t)=0$ for $\gamma>\gamma_{\rm cr}$ for the shear acceleration analysis in the 
following.
Thus our solution only applies for particles with $\gamma>\gamma_{\rm cr}$, and is entirely independent of the particle distribution below $\gamma_{\rm cr}$.

As the dynamical timescale is larger than the acceleration timescale 
for large-scale jets and even for individual knots (see Eqs. (\ref{tdyn}-\ref{chi}) in Section \ref{section:ordersofmagnitude}), 
%Further, as the observed X-rays are quasi-continuous along the jet, 
it is appropriate to take a quasi-steady-state approximation, i.e. ${\partial n(\gamma,t)/\partial t}=0$. 
In this case, we find that Eq. (\ref{eq:Fokker}) has a general solution. 
The solution for $q=1$ is given in Eq. (\ref{solq=1}). For $1<q\leq2$, which is of more common astrophysical 
interest, the solution is (see Appendix \ref{sec:Solq} for a derivation), 
\begin{equation}
\label{solq}
n(\gamma)= C_+\gamma^{s_+} F_+(\gamma, q)  
+ C_-\gamma^{s_-} F_-(\gamma, q), %\left({2\over q-1}{A_2\over A_1}\right)^{{s_+\over q-1}}\left({2\over q-1}{A_2\over A_1}\right)^{{s_-\over q-1}}
\end{equation}
where 
%$C_\pm$ are constants which are not independent, as they are related by the requirement $n\to0$ for $\gamma\to\infty$  (see Appendix \ref{sec:Solq}). The remaining constant determines the normalization of the spectrum.
the power-law spectral indices are 
\begin{equation}
s_\pm={q-1\over2}\pm\sqrt{{(5-q)^2\over4}+w},\label{eq:spm}%(6-q){\tau_{\rm sh}\over\tau_{\rm esc}}=\left\langle {10 c^2\over \Gamma_{\rm j}^4 \Delta v_{\rm j}}\right\rangle 
\end{equation}
where 
\beq
w={(6-q)\tau_{\rm sh}\over \tau_{\rm esc}}
%2/(A_1A_3)
={4 c^2\over 3 D_{\rm sh}\drj^2}\label{eq:w}
\eeq
is a dimensionless measure of the shear viscosity.
This indicates that the particle spectral hardness is mainly determined by the turbulence spectrum and the balance of acceleration and escape. 
For $1<q<2$ the indices satisfy $s_+>0$ and $s_-<0$. 

The $F_\pm$ are defined as
\begin{equation}
F_\pm(\gamma, q)={_1F_1}\left[ {2+s_\pm \over q-1},{2s_\pm\over q-1};-{6-q\over q-1}\left({\gamma\over \gmax}\right)^{q-1}\right],
\end{equation} 
where $_1F_1$ is Kummer's confluent hypergeometric function \citep[e.g.,][]{Abram1972}.
For negligible radiative losses, i.e. for $\gamma\ll\gmax$, $F_\pm\approx 1$, such that the solution resembles a power-law. The power-law index is consistent with previous results \citep{Kimura2018,Rieger2019ApJ}. 

The spectrum is obtained from the combination of two power-law components with the integration constants ($C_\pm$) being determined by two requirements, i.e., the condition of $n\to0$ for $\gamma\to\infty$ (see Appendix \ref{sec:Solq}) and the normalization of the spectrum. We find that the spectrum exhibits an exponential-like-cutoff power-law shape, where the power-law index is dominated by the $s_-$ component. 
 
%By numerically comparison with the exponential function (exp$[-(\gamma/\gmax)^{y}]$), we find the cut-off can be slightly super-exponential with $y\sim1-2$.

\begin{figure}
	\centering
	\includegraphics[width=0.49\textwidth]{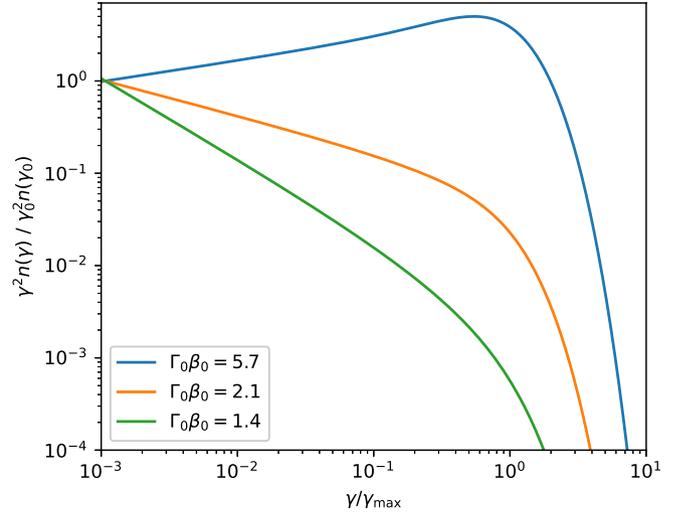}
	\caption{Three examples of particle energy distributions are calculated from Eqs. (\ref{solq}-\ref{eq:wlinear}) with different velocities, i.e. $\Gamma_0\beta_0=1.4,~2.1,~5.7$, where $q=5/3$ is adopted.
	The spectra are normalized relative to their value at $\gamma_0=10^{-3}\gmax$.
	}
	\label{fig:1}
\end{figure}

The particle spectral index can be specifically determined for a given
shearing profile. 
Here we consider a simple shearing profile with a jet speed linearly
decreasing from $\beta_{\rm j}=\beta_0$ at $r_{\rm j}-\drj$ to $\beta_{\rm
j}=0$ at $r_{\rm j}$, where $r_{\rm j}\geq\drj$ is the jet width. Performing
a simple r-averaging, one obtains  \citep[see Eq. (9) in][]{Rieger2019ApJ}
\beq
w=40\ln^{-2}{(1+\beta_0)\over(1-\beta_0)}.\label{eq:wlinear}
\eeq
For this profile and assuming $q=5/3$, we show three examples of the 
resulting particle distributions in Figure \ref{fig:1}. As one can see, 
the spectral shape is highly dependent on the jet speed $\beta_0$.
In the highly relativistic limit, we have $w\rightarrow 0$, and the spectrum with index $s_- = q-3$ is approached \citep[e.g.,][]{Webb2018,Webb2019,Rieger2019ApJ}, in which particle escape will be insignificant, i.e. $\tau_{\rm esc}\gg\tau_{\rm sh}$. 
While for non-relativistic flow speeds, steep spectra are obtained. 
In general, the spectral index is sensitive to the 
employed velocity profile, with somewhat steeper shapes towards lower
speeds being possible \citep{Webb2019}, while harder spectra may become
possible if particles could re-enter the jet again \citep{Webb2020}.
%\beq\label{beta:linear}
%\beta_{\rm j}(r) =\begin{cases}
%	\beta_0,~\text{for}~r<\zeta r_{\rm j},\\ 
%	\beta_0(1-r/r_{\rm j})/(1-\zeta),~\text{otherwise.}
%\end{cases}
%\eeq
%where $r_{\rm j}$ is the jet radius, and the size of shearing region is $\drj=(1-\zeta)r_{\rm j}$ with $\zeta\in[0,1]$. 
%With this profile, we then average Eq. (\ref{Diffu_shear}) over $r$ to obtain the mean shear coefficient $\left<D_{\rm sh}\right>=\int_{0}^{r_{\rm j}}2\pi r D_{\rm sh}dr/\int_{0}^{r_{\rm j}}2\pi r dr$.
%Combining Eq. (\ref{beta:linear}), we obtain 
%\begin{eqnarray}
%w&=&{20\,(1-\beta_0^2)(1-\zeta)\over \beta_0[\beta_0\zeta+(1-\beta_0^2){\rm Arctanh}(\beta_0)]}.\label{eq:wlinear}
%w_{\rm linear}&=&{10\over \beta_0 {\rm arctanh}(\beta_0)};\\
%w_{\rm Gaussian}&=&{10(\exp[2a]-\beta_0^2)\over (\exp[2a]-\beta_0^2){\rm ln}[\Gamma_0^2(\exp[2a]-\beta_0^2)]-2a\exp[2a]}.
%D_{\rm sh}&=& {2c^2\beta_0 {\rm arctanh}(\beta_0)\over15\Delta r_{\rm j}^2},\\
%s_- &=& -\sqrt{{(5-q)^2\over4}+{10\over \beta_0 {\rm arctanh}(\beta_0)}}+{q-1\over2}.
%\end{eqnarray}

%\begin{figure}
%	\centering
%	\includegraphics[width=0.49\textwidth]{fig1.eps}
%	\caption{The particle spectral index parameter $w$ and $s_-$ as functions of $\Gamma_0\beta_0$ are obtained assuming a Kolmogorov-type turbulence ($q=5/3$). Three cases with different linear profiles are studied, i.e. $\zeta=0.0,~0.2,~0.8$. We also show the power-law index from previous work \citep{Webb2018} for comparison. }
%\end{figure}

\section{Radiation from accelerated electrons in large-scale shearing jets}\label{section:X-ray-jet}

\subsection{General radiation features}\label{section:ordersofmagnitude}

In the following we examine the capability of particle acceleration via the shear mechanism, and study 
its associated radiation features \citep[cf. also][]{Liu2017,Rieger2019ApJ}. As shown above, in
steady-state the accelerated particles produce power-law spectra with an exponential-like cut-off. Such a steady 
state can only be achieved when the dynamical time ($\td$) greatly exceeds the acceleration time. 
For trans-relativistic jets, the dynamical time can be expressed as 
\begin{equation}
    \td=L_z/\beta_0c =3.26\times10^2\chi \rjc\beta_0^{-1}~
{\rm yrs},\label{tdyn}
\end{equation}
where the length of the acceleration region is taken to be $L_z=\chi r_{\rm j}$, with $r_{\rm j}=10^2\rjc$ pc. 
We set $\Lambda_{\rm max}=\xi_2 \drj $ such that the outer scale of the turbulence is less than the characteristic shearing scale-width. 
We define $\eta=\xi_2^{q-1}\xi_1^{-1}$, and from this point forward restrict our focus to Kolmogorov turbulence $q=5/3$, other cases being a trivial extension. 
The acceleration time is given numerically as 
\begin{equation}
    \tau_{\rm sh}=1.38\times10^{3} w \bc^{1/3}\drjc^{4/3}\gf^{-1/3}\eta^{-1}~
{\rm yrs},\label{tacc_e}
\end{equation}
where $\gamma=10^9\gf$. We find that the condition for the application of the steady-state approximation is met when the ratio of jet radius to acceleration length satisfies
\begin{equation}
    \chi\gtrsim4.2w\beta_0 \bc^{1/3}\drjc^{4/3}\rjc^{-1}\gf^{-1/3}\eta^{-1}.\label{chi}
\end{equation}
For large-scale AGN jets, this condition is likely to be satisfied.
%For observed jets with continuous X-ray emission, particle acceleration 
%is also expected along the jet; and this criterion can be generally satisfied.
%%Observations of collimated jets confirm that this criterion is generally satisfied.
Even in jets where the X-ray emission is dominated by knotted structures, this 
criterion might be met on the scale of individual knots, at least marginally.
%While X-ray jets are often dominated by knotted structures, we now show that 
%even for individual knots this criterion can also be satisfied, at least marginally.
Observations suggest that for typical knots $\chi>2/\sin{\theta}$, where $\theta$ 
is the viewing angle to the jet axis, which can be small. For example, in the
case of 3C 273, $\theta\lesssim 0.13$ has been reported \citep[e.g.][]{Meyer2016}, 
which would lead to $\chi>16$.
%The physical jet length will naturally be longer than that inferred from observations due to projection effects. 

The scattering timescale may also be expressed numerically as
\begin{equation}
    \tau_{\rm sc}=26.8 \drjc^{2/3} \gf^{1/3}\eta\bc^{-1/3}~
{\rm yrs}. 
\end{equation}
Here, to ensure particles are in fact accelerated, we require that at least one scattering time can occur, i.e. $\tsc\lesssim\tsh$. This leads to an additional constraint
\begin{equation}
    \eta\lesssim 7.17w^{1/2}\bc^{1/3}\drjc^{1/3}\gf^{-1/3}. \label{eta1}
\end{equation}

The accelerated electrons suffer from radiation losses on a characteristic cooling time
\begin{equation}
    \tau_{\rm c}=2.45\times 10^{2}\bc^{-2} (1+f)^{-1}\gf^{-1}~
{\rm yrs}.
\end{equation}
Acceleration proceeds provided $\tsh \lesssim \tau_{\rm c}$, which translates to another condition
\begin{equation}
\eta\gtrsim 5.61w\bc^{7/3}(1+f)\drjc^{4/3}\gf^{2/3}.\label{eta2}
\end{equation}

The inequalities (\ref{eta1}) and (\ref{eta2}) further require $\gf\lesssim1.28\bc^{-2}(1+f)^{-1}
\drjc^{-1}w^{-1/2}$, while the Larmor radius is $r_{{\rm L}, e}= 0.05\gf\bc^{-1}$ pc. 
Thus the Hillas criterion can be satisfied, and the corresponding maximum energy for electrons is 
\beq
E_{e,{\rm max}}=0.7\bc^{-2}\drjc^{-1}w^{-1/2}(1+f)^{-1}~{\rm PeV},\label{eq:true_gmax}
\eeq
The resultant maximum synchrotron photon energy that the electrons may therefore radiate is 
\beq
E_{\rm \gamma, max}=82.3\bc^{-3}\drjc^{-2}w^{-1}(1+f)^{-2} ~{\rm keV}.\label{Emaxsyn}
\eeq
This demonstrates that electrons, energized via shear acceleration can produce X-rays. In large-scale jets, such X-rays are naturally expected to be emitted quasi-continuously along the jet. We now apply this model to selected observations of X-ray jets.

\subsection{Application to large-scale X-ray jets\label{sec:application}}

Observations show that large-scale jets can be bright in X-rays even on scales of one-hundred-kpc \citep[e.g.][]{Harris2006}. 
Although the mechanism producing this X-ray radiation is still under debate, a synchrotron origin is strongly supported by current evidence e.g. the extended TeV radiation from Centaurus A \citep{HESS2020}, 
and the optical polarimetry and gamma-ray observations of 3C 273 \citep{Meyer2014,Perlman2020}. 
In the following, we take these two jets as examples. 
We show that their X-ray signatures can be explained in the framework of shear acceleration with typical jet parameters. 
Note the fitting parameters are exemplary, not unique. 
The SED modelling is performed using the NAIMA package \citep{naima}. This package includes tools 
to calculate the non-thermal radiation from relativistic particle populations and to perform
Markov-Chain-Monte-Carlo fitting of the observed spectra. % using emcee package \citep{emcee}.
Here, we employ NAIMA to calculate the (one-zone) synchrotron and IC emission, and to model the low-energy electron distribution, which is treated as independent input and assumed to be a (cut-off) power-law distribution for FR I (II) sources; the high-energy electron distribution is modelled with Eqs. (\ref{solq}-\ref{eq:wlinear}, \ref{eq:true_gmax}).

\subsubsection{FR I jet: Centaurus A}
Centaurus A (NGC 5128) is an FR~I radio galaxy located at a distance of $3.8$ Mpc \citep{Harris2010}. 
Radio observations of the jet's proper motion on the sub-kpc scale suggest that the jet is moving with a mildly relativistic speed of $\beta_0\sim0.5$ \citep{Hardcastle2003,Snios2019}.
Recently, the High Energy Stereoscopic System (H.E.S.S.) has detected extended very-high-energy (VHE) emission from the large-scale jet of Centaurus A \citep{HESS2020}.  

\begin{figure}
    \centering
    \includegraphics[width=0.49\textwidth]{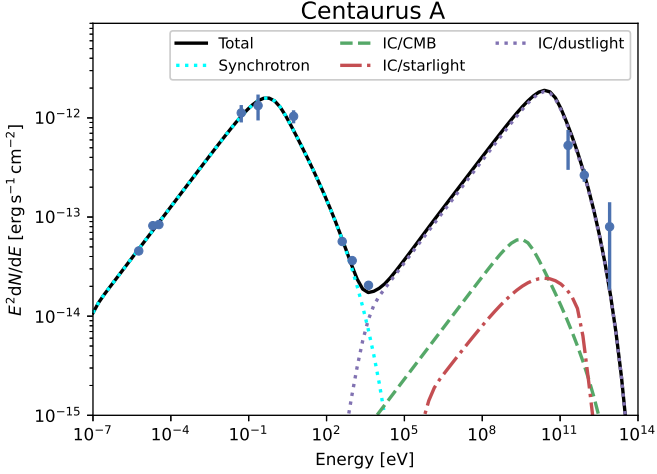}
    \caption{The multi-wavelength SED of Centaurus A reproduced by synchrotron and IC radiation of 
    electrons with an exponential-cut-off broken power-law spectrum in the framework of shear acceleration. 
    The model parameters are shown in Table \ref{tab:fitting}. }
    \label{fig:2}
\end{figure}

In Figure \ref{fig:2}, we show an exemplary multi-wavelength-SED fit of the large-scale jet emission of Centaurus A. The data are taken from \citet{Hardcastle2006} and \citet{HESS2020}. The electrons 
are assumed to follow a broken power-law spectrum with an exponential cut-off. We find that 
the radio-to-optical SED can be explained by synchrotron radiation from electrons with $N_1(E)= K_1 E^{-\alpha_1}$ for $E_{\rm min,1}\leq E\leq E_{\rm b}$ and $K_1$ being the normalization constant. 
The X-ray and TeV data, on the other hand, is modelled as synchrotron and IC radiation from shear 
accelerated electrons ($N_2(E)$), i.e. for $E\geq E_{\rm b}$, we adopt Eqs. (\ref{solq}-\ref{eq:wlinear}, \ref{eq:true_gmax}), and the normalization is 
set to keep $N_1(E_{\rm b})=N_2(E_{\rm b})$. 
The model has in total 7 free parameters: $K_1,~E_{\rm min,1},~E_{\rm b},~\alpha_1,~w,~B,~\drj$. 
For the IC spectrum, the seed photons come from the CMB, starlight \citep{Abdo2010}, and the radiation 
from dust \citep{Weiss2008}. 
The seed photon energy density is found to be much smaller than that of the magnetic field, i.e. $f\ll1$. 
%The TeV SED is consistent with IC radiation from a population of electrons ($dN/dE\propto E^{-\alpha_h}$) with a spectral index $\alpha_h\sim3.6$ \citep{HESS2020}. 

The model parameters are shown in Table \ref{tab:fitting}. 
The radio-to-optical fit is consistent with previous findings \citep{Hardcastle2006,HESS2020}. 
The X-ray and TeV spectra can be reproduced with $w=15.0$, $B=17.1~\mu$G, and $\drjc=1$ for our
shear-acceleration model. This implies a spectral index of $s_-=-3.9$, a jet velocity at $\beta_0\approx0.67$, and an electron energy cut-off at $57$ TeV. %The corresponding maximum synchrotron photon energy is $E_{\rm \gamma, max}\approx6\bc^{-3}\drjc^{-2}(1+f)^{-2} ~{\rm keV}$ from Eq. (\ref{Emaxsyn}). 
%The minimum electron energy is assumed to be $E_{\rm min, e}=0.1$ GeV, and the break energy is $E_{\rm b}=0.8$ TeV. 
%For the X-ray and TeV radiation, the electrons are assumed to be re-accelerated due to shear acceleration, with fitting parameters $\beta_0\sim0.68$ and $B=20\drjc^{-1/2}~\mu$G, implying a spectral index $s_-=-3.8$ and an electron cut-off energy to be $40$ TeV. 
The (spine) velocity is slightly larger than that estimated based on radio observations.
This difference is reasonable given the uncertainties in the shearing profile and the fact that the radio/optical emission may not fully trace the inner spine speed.
%The total electron energy in this case is $2.7\times 10^{53}$ ergs, compatible with the findings reported by the H.E.S.S. Collaboration\citep{HESS2020}.
From the morphology study in radio and X-ray band, the projected jet length is found to be a few kpc, and 
the half jet width can be of order $\drjc\sim1$ \citep{Hardcastle2003}. Thus both the X-ray and TeV radiation can 
in theory be accounted for exclusively within a shear-acceleration model. 
The kinetic jet power can be estimated as $P_{\rm jet}=(W_e+W_p+W_B)\beta c/L_z$ where $W_e,~W_p,~W_B$ is the total energies of relativistic electrons, cold protons, and magnetic field in the radiating region. %This is justified as the total energy in electrons exceeds that of both, the magnetic field and protons. The latter follows from the observation that the inferred minimum electron energy is larger than the proton rest energy, i.e. $E_{\rm min,1}>$ GeV, since smaller values would over-produce X-ray fluxes by IC scattering. 
Taking the jet length to be $L_z\sim2$ kpc, we obtain $P_{\rm jet}\sim3.7\times10^{42}$ erg/s for Cen A. 
This is much smaller than the Eddington luminosity of Cen A, which is $L_{\rm Edd}\sim 6.9\times 10^{45}$
erg/s for a mass $5.5\times 10^7M_\odot$ \citep{Cappellari2009}, and compatible with constraints on the average jet power inferred from modelling of the lobes in Cen A \citep{Yang2012A}.

\subsubsection{FR II jet: 3C 273}
\begin{figure*}
    \centering
    \includegraphics[width=0.49\textwidth]{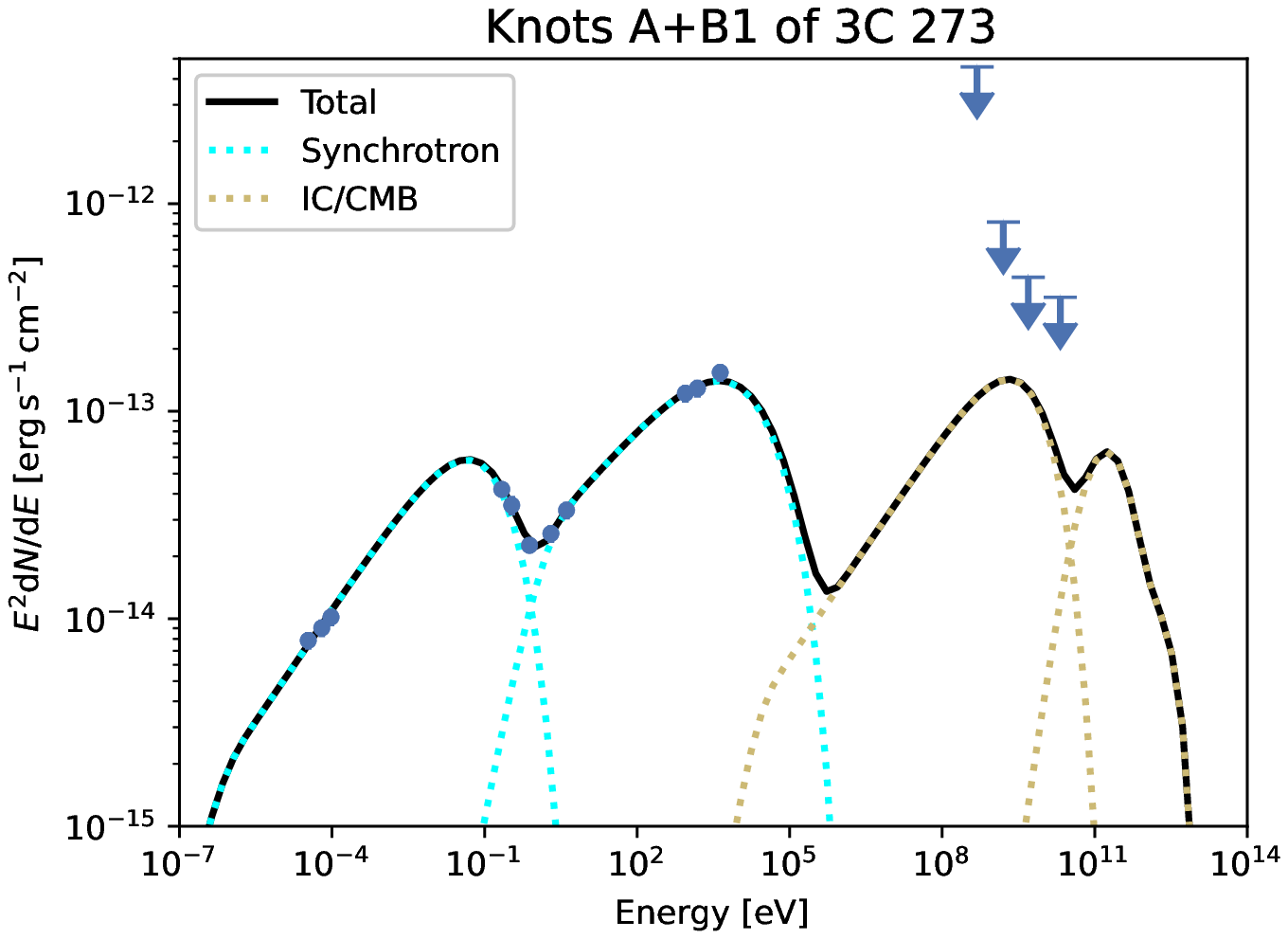}
    \includegraphics[width=0.49\textwidth]{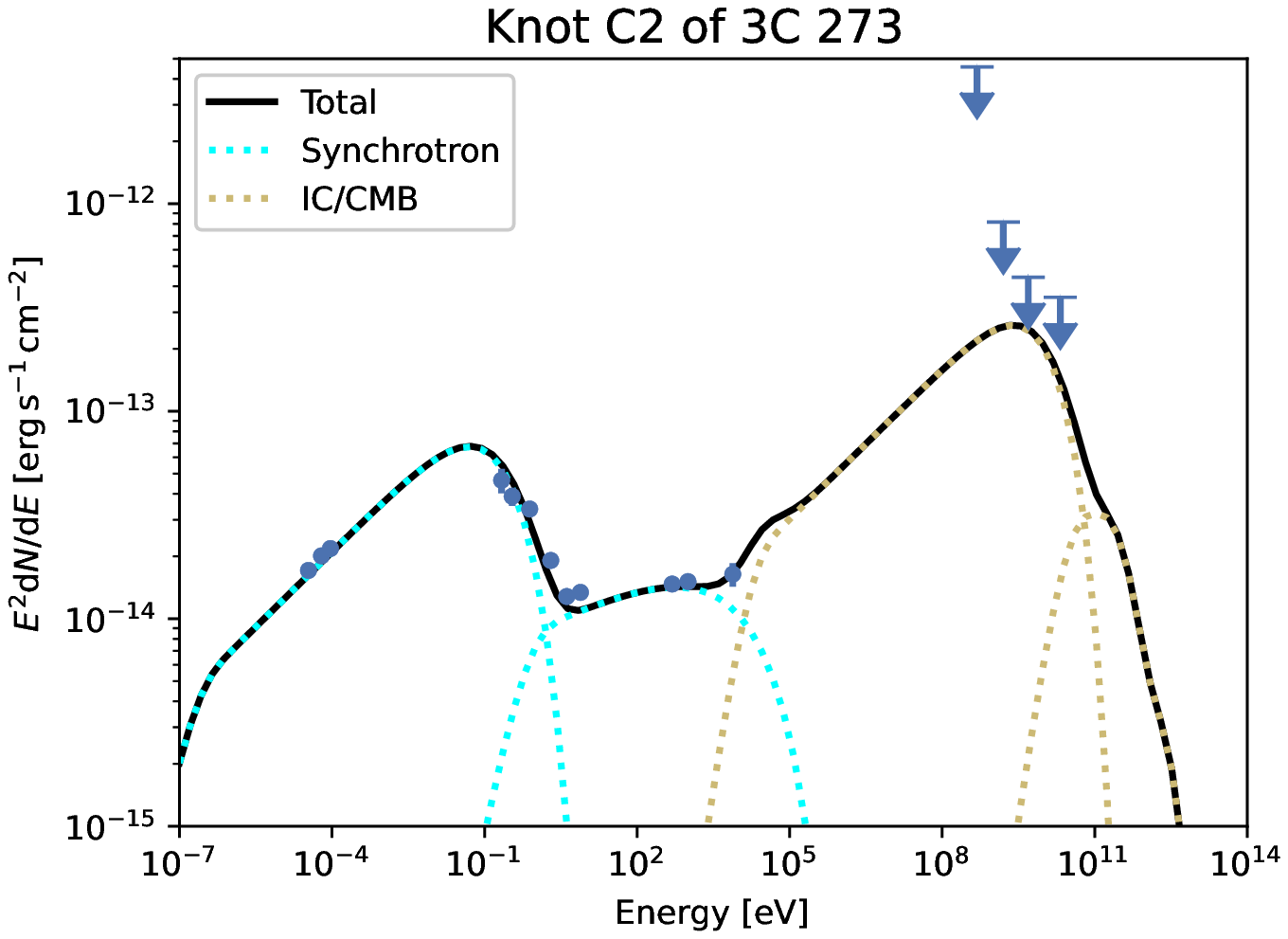}
    \caption{The multi-wavelength SED of the knots in 3C 273 (left panel for Knots A+B1 and right panel 
    for knot C2) reproduced by synchrotron and IC radiation of two electron populations with exponential-cut-off power-law spectra in the framework of shear acceleration. 
    The model parameters are shown in Table \ref{tab:fitting}. }
    \label{fig:3}
\end{figure*}

3C 273 is an FR II radio galaxy, located at a redshift of $z=0.158$. 
In general, FR II jets are dominated by bright knotted structures, which often exhibit very hard X-ray spectra with X-ray spectral indices comparable to their radio spectral indices. For 3C 273 only the first two knots have such hard X-ray spectra, while the X-ray spectra become softer 
for outer knots \citep[][see also our Figure \ref{fig:3}]{Jester2006}. 
We here take the first two knots (A and B1) and an outer knot (C2) as examples. Below we show that the primary spectral features in the three 
knots can be naturally accounted for in the shear acceleration model. 
The radio, optical, and X-ray data are taken from \citet{Jester2007}. 
The $\gamma$-ray upper limits are from \citet{Meyer2014}.

In Figure \ref{fig:3}, we show exemplary fits of knots A and B1, and C2, where the multi-wavelength SEDs 
are well reproduced. 
The radio-to-optical SED is explained by synchrotron radiation from electrons with $N_1(E)=K_1 E^{-\alpha_1}\exp[-(E/E_{\rm b})^2]$ for $E>E_{\rm min,1}$, where $K_1$ is the normalization constant, 
and the super exponential cut-off follows from cooling effects \citep{Zirakashvili2007}. The X-ray SED 
is explained by synchrotron emission from shear accelerated electrons ($N_2(E)$), which follows Eqs. (\ref{solq}-\ref{eq:wlinear}, \ref{eq:true_gmax}) with a normalization constant $K_2(C_+,C_-)$ for 
$E>E_{\rm min,2}$. Thus, two additional free parameters are necessary, i.e. we have $K_1,~E_{\rm min,1},
~E_{\rm b},~\alpha_1,~K_2,~E_{\rm min,2},~w,~B,~\drj$.
The IC/CMB radiation for the two electron populations are calculated with the extragalactic 
background light (EBL) absorption taken into account following \citet{EBL2011}. 

The model parameters are shown in Table \ref{tab:fitting}. 
The radio-to-optical SEDs are consistent with a synchrotron interpretation as suggested by previous 
analysis \citep{Jester2007}. The UV-to-X-ray SED can be matched using $\drjc=10$, $w=4.7$, $B=2.8~\mu$G 
($w=6.8$, $B=2.2~\mu$G) for knots A+B1 (C2) within the shear acceleration framework. 
This implies a spectral index of $s_-=-2.4$ ($-2.8$), a jet velocity at $\beta_0\approx0.90$ ($0.84$), 
and an electron-energy cut-off at $370$ ($490$) TeV for knots A+B1 (C2).
The jet speed is consistent with the upper-limit from proper-motion studies, which suggest $\Gamma_0<2.9$ 
for the knot bulk Lorentz factor \citep{Meyer2016}. 
From the X-ray morphology, the jet width is $r_{\rm j}\sim 1$ kpc \citep{Jester2006}, and the 
length is $\sim4/\sin\theta$ ($3/\sin\theta$) for knots A+B1 (C2), where $\theta\sim0.13$
\citep[e.g.,][]{Meyer2016}. 
The corresponding jet power is $2.7\times 10^{45}$~ ($1.3\times 10^{46}$) erg/s for knots A+B1 (C2).
This is much smaller than the Eddington luminosity of 3C 273, $L_{\rm Edd} \sim 8.2\times 10^{47}$, 
adopting a black hole mass of $6.6\times 10^9M_\odot$ \citep{Paltani2005}.
%The radio to optical SED is fitted by synchrotron radiation from a population of electrons with a spectral index $\alpha\approx2.3$ and a cut-off at $\sim1$ TeV, consistent with previous findings \citep{Jester2006}. 
%The UV to X-ray SED is fitted by the synchrotron radiation from a distinct population of high-energy electrons with a spectral index $\alpha\approx2.4$ and minimum and cut-off energies at $\sim2.5$ TeV and $\sim370$ TeV, respectively. 

\begin{table*}
    %\centering
    \caption{The exemplary parameters for the fitted SEDs of large-scale jets, Centaurus A and 3C 273. The low-energy component are responsible for radio-to-optical SED, while the high-energy component are responsible for the X-ray SED. 
    The jet speed is calculated from Eq. (\ref{eq:wlinear}). }
    \begin{tabular}{c|c|c|c|c}
        \hline
        Source & Low-energy component & High-energy component & Jet power in units of & Jet \\  
        name & $(E_{\rm min,1},~E_{\rm b},~\alpha_1)$ & ($E_{\rm min,2}~,w,~B,~\drjc$) & erg/s and $L_{\rm Edd}$ & speed \\ \hline
        Centaurus A & $0.2$ GeV, $0.75$ TeV, $-2.31$ & -, $15.0$, $17.1~\mu$G, $1$  & $3.7\times10^{42}$ erg/s, $5.4\times10^{-4}L_{\rm Edd, Cen A}$ & $0.67c$\\\hline
        3C 273 - Knots A+B1 & $1.5$ GeV, $1.1$ TeV, $-2.28$ & $2.5$ TeV, $4.7$, $2.8~\mu$G, $10$ & $2.7\times 10^{45}$ erg/s, $3.2\times10^{-3}L_{\rm Edd, 3C 273}$ &$0.90c$\\\hline
        3C 273 - Knot C2 & $1.5$ GeV, $1.6$ TeV, $-2.52$ & $1.9$ TeV, $6.8$, $2.2~\mu$G, $10$ & $1.3\times 10^{46}$ erg/s, $1.5\times10^{-2}L_{\rm Edd, 3C 273}$ & $0.84c$\\\hline
    \end{tabular}
    %{\begin{flushleft}\end{flushleft}}
    \label{tab:fitting}
\end{table*}

\section{Implications for UHECR acceleration}\label{section:CR}

AGN jets are proposed as potential ultra-high-energy cosmic-ray (UHECR) accelerators based on the Hillas criterion
\citep{Hillas1984,Aharonian2002}, although the acceleration mechanism(s) that might achieve such energies is an open question.
As shown above, large-scale jets can in principle accelerate electrons up to PeV energies. We now study its potential for CR acceleration. 
To ensure that particles can be accelerated in trans-relativistic jets, the condition $\td>\tsh$ must be satisfied, which requires seed nuclei to have energies
\begin{equation}
E_{{\rm seed~nuclei}}>38 Z\bc \drjc^4 w^3\beta_0^3\chi_1^{-3}\eta^{-3}\rjc^{-3}~{\rm TeV},
    %\chi>0.34w\beta_0 \bc^{1/3}\rjc^{1/3}\drjc^{-2/3}\gf^{-1/3}\eta^{-1},
\end{equation}
where $\chi_1=\chi/10$, $Z$ is the atomic number, and $\eta\sim 1$ encapsulates the turbulent magnetic field strength and scales defined previously.

The maximum attainable energy is limited by several requirements. For consistency, we require $\tsc\lesssim\tsh$, form which we obtain 
\begin{equation}
    E_{\rm nuclei}\lesssim 0.2Z\bc\drjc w^{3/2}\eta^{-3}~\rm{EeV}.
    \label{eq:gmax_p}
\end{equation}
Cooling via synchrotron radiation is irrelevant for the fields we consider.
%, as $\tsh\lesssim\tsy$ can be satisfied for $E_{p}\lesssim 10^{15}\bc^{-7/2}\drjc^{-2} w^{-3/2}\eta^{3/2}~\rm{EeV}$.
%\begin{equation}
    %E_{p,{\rm max}}=7.4\bc^{-2}\drjc^{-1}w^{-1/2}(1+f)^{-1}~\rm{EeV}.\label{eq:gmax_p}
%\end{equation}
We further note that the Larmor radius of a nucleus is $r_{{\rm L}}=0.1 \gf\bc^{-1}Z^{-1}~{\rm kpc},$
which can be comparable to the size of the shearing region. Based on the Hillas criterion, we restrict the 
Larmor radius to be inside this region, i.e., %. As such, a particle with charge $Ze$ should satisfy
\begin{equation}
    E_{\rm Hillas}\leq0.9 Z \beta_0 \bc\drjc~{\rm EeV}.
\end{equation}
Thus the maximum energy may therefore be expressed as
\begin{equation}
    E_{{\rm max}}= {\rm min}\left[0.9,~0.2 w^{3/2}\eta^{-3}\right]~ Z \bc\drjc ~{\rm EeV}. \label{eq:CR_max}
\end{equation}
For large-scale AGN jets, EeV protons are in theory achievable, while heavier nuclei would extend to higher energies. 
Note that for EeV protons, the characteristic synchrotron photon energy is $E_{p,{\rm syn}}\approx 30\gf^2\bc$ eV \citep{Aharonian2002proton}. We conclude that shear acceleration is a promising mechanism for the acceleration of UHECRs in large-scale AGN jets \citep[e.g.,][]{Liu2017, Kimura2018, Webb2018,Rieger2019ApJ}.

%Observations of UHECRs suggest a trend toward heavier compositions above $\sim4$ EeV \citep[see][for a review]{BeattyWesterhoff}. If we assume that shear acceleration produces nuclei right up to the Hillas limit, then protons can be accelerated to $10$ EeV, while nuclei with charge $eZ$ might reach a factor of $Z$ times greater energies, e.g. $260$ EeV for iron. Accelerating protons to $10^{20}$ eV would require $\bc\drjc\approx10^2$ from Eq. (\ref{eq:CR_max}). Substituting this to Eq. (\ref{Emaxsyn}), we obtain $E_{\rm \gamma, max}=0.1\drjc w^{-1}(1+f)^{-2} ~{\rm keV}$.Therefore, acceleration of highest-energy CRs is more favourable in the location with softer X-ray SED, which is usually far-away from the AGN core. Overall, 

The accelerated proton spectrum within the jet should approximately follow Eq. (\ref{solq}).  
On the other hand, the spectrum of the escaping protons, which can contribute to the observed CRs, follows 
\begin{equation}
    \dot{n}_{\rm esc}\propto {\gamma^{s_-}/\tes}=\gamma^{2-q+s_-}.
\end{equation}
This spectrum is harder than the confined particle spectrum. The X-ray observations suggest spectral indices of $|s_-|\approx3.6$ for Centaurus A and $2.4(2.8)$ for 3C 273, such that the implied index for escaping protons is $3.3$ for Centaurus A and $2.1(2.5)$ for 3C 273.

\section{Summary and Conclusion}\label{section:conclusion}

In this paper, we have explored the capability of gradual shear flow particle acceleration in large-scale X-ray jets. 
Velocity-shear stratification is naturally expected in large-scale jets, and energetic seed particles can be accelerated by interacting with the magnetic field inhomogeneities frozen in stratified layers. The seed particles can be injected from shock acceleration, magnetic reconnection, and/or stochastic acceleration.
We provide a general solution of the steady-state Fokker-Planck equation for shear acceleration in trans-relativistic jets. In general, the accelerated particle spectrum resembles an exponential-like-cutoff-power-law spectrum. The power-law index is essentially determined by the turbulence spectrum, and the balance of escape and acceleration of particles (see Eq. 
\ref{eq:spm}). %The cutoff shape (exp$[-(\gamma/\gmax)^{y}]$) is found to be $y\sim1-2$. 
Assuming a simple linearly decreasing velocity profile in the boundary of large-scale jets, the particle spectrum can be determined (Eq. \ref{eq:wlinear}). The maximum energy of the electrons is typically found to be $\sim$ PeV, radiating $\sim$ keV X-rays via synchrotron radiation. Such synchrotron origin of the X-ray emission in knots of powerful jets has also recently been investigated by \citet{Tavecchio2020}, where shear acceleration of a locally produced shock-accelerated particle population is explored numerically. %While the distributed acceleration process motivating our study is conceptually different, the formalism is mathematically similar, and in fact the numerical results of \citet{Tavecchio2020} and our steady-state result should agree in equilibrium.

We have applied our model to observations of the X-ray jets in Centaurus A and 3C 273. 
In FR I jets, the radio-optical-X-ray spectrum usually conforms to synchrotron radiation from electrons with a broken-power-law spectrum. 
For Centaurus A, the recent TeV observations further support the synchrotron origin of X-rays. 
In FR II jets, the optical polarimetry and $\gamma$-ray observations disfavor the popular IC origin of X-rays, while electron or proton synchrotron remain viable. 
We note however that proton synchrotron in X-rays usually requires $\sim$mG magnetic fields in kpc-scale jets \citep{Aharonian2002proton,Wang2020}. 
Modelling of the radiative features \citep[e.g.][]{Sikora2005,Potter2017} as well as MHD simulations \citep[e.g.][]{Chatterjee2019} both suggest that the jet is kinetic energy dominated on such scales. 
Therefore, an electron synchrotron origin of the X-ray emission is favored for both FR I and II jets. 
To explain the extended X-ray emission at kpc-scale jets, an {\em in-situ} (re-)acceleration mechanism is required due to the significant synchrotron radiation cooling. 
In contrast to shock acceleration, which would result in the most energetic electrons being localised to the position of a shock, and magnetic reconnection or related processes, which are only effective in magnetically dominated jets at $\lesssim$ pc scale \citep{Matthewsetal}, we find shear acceleration can naturally accelerate particles along the jet even at $>$ kpc scales. 
In our shear-acceleration model, the multi-wavelength SED (Figures \ref{fig:2} and \ref{fig:3}) can be satisfactorily reproduced by synchrotron and IC radiation with typical jet parameters (Table \ref{tab:fitting}) 
for Centaurus A and 3C 273. 
The required jet power in our exemplary fittings is significantly below the respective Eddington Luminosity of each source, with $P_{\rm jet}\sim4\times10^{42}$ erg/s %$P_{\rm jet}/L_{\rm Edd}\sim5\times10^{-4} $ 
for Centaurus A and $P_{\rm jet}\sim10^{45}-10^{46}$ erg/s %$P_{\rm jet}/L_{\rm Edd}\sim(10^{-3}-10^{-2})$ 
for 3C 273. This, as well as the derived jet velocities at $\sim0.67c$ for Centaurus A and $\sim0.9c~(0.84c)$ for the knots A+B1 (C2) of 3C 273, is consistent with the general picture of FR II jets being more powerful than FR I jets. 
Further, the decelerating jet of 3C 273, which in our model should decrease from $\sim0.9c$ at knots A+B1 to $\sim0.84c$ at knot C2, is also naturally expected as the jet propagates. 
We note that two populations of electrons are required in 3C 273, which may hint that the radio-to-optical and X-ray SEDs are produced in different locations, for example in the spine and sheath picture \citep[e.g.,][]{Jester2006}. However, only one population of electrons is needed in Centaurus A. This difference may relate to jet dynamics, such as the properties of the shearing region, and should be followed up in future work.

For shear acceleration in large-scale jets, protons and other nuclei can also be accelerated with a spectrum similar to that of electrons, while the escaping particles could have an even harder spectrum. The maximum energy is found to be $E_{{\rm max}}={\rm min}[0.9,~0.2 w^{3/2}\eta^{-3}] Z\bc\drjc$~EeV, consistent with previous findings \citep{Liu2017,Rieger2019ApJ}. 
This work further supports the evidence that shear acceleration can provide a favourable mechanism for UHECR acceleration in large-scale jets.

\section*{Acknowledgements}
We thank the referee for valuable comments, and X. N. Sun, Y. Mizuno and M. Tsirou for helpful discussions. 
J.S.W. was supported by China Postdoctoral Science Foundation. R.-Y. L. is supported by NSFC No.~U2031105.

\section*{Data Availability}
%There are no new data associated with this article.
No new data were generated or analysed in support of this research.

\bibliographystyle{mnras}
\bibliography{ref}

\onecolumn 
\appendix

\section{Derivation of steady-state solution for Equation (8)}\label{sec:Solq}
We here present a brief derivation of Eq. (\ref{solq}). 
Starting from Eq. (\ref{eq:Fokker}) we seek a steady-state (${\partial n(\gamma,t)/\partial t}=0$) solution 
above some minimum injection energy, i.e. $Q(\gamma>\gamma_{\rm cr},t)=0$ where $\gamma_{\rm cr}$ is the 
maximum energy particles injected into the jet.
We seek solutions of the form $n(\gamma)\propto \gamma^s j(\gamma)$.
Using $w=-6+2q+(1-q+s_\pm)\,s_\pm$ and setting $n(\gamma)\propto \gamma^{s_\pm} j_\pm(\gamma)$, 
Eq. (\ref{eq:Fokker}) can be reduced to 
\begin{align}
&2(A_2/A_1)(s_\pm+2)\gamma^{q-3}j_\pm(\gamma)+ 
[\gamma^{-1}(2-q+2s_\pm)+2(A_2/A_1)\gamma^{q-2}]j_\pm'(\gamma)+j_\pm''(\gamma)=0.
\end{align}

For $1<q\leq 2$, we can introduce a new variable $z= -\frac{6-q}{q-1} (\gamma/\gamma_{\rm max})^{q-1}$ and its inverse $\gamma(z)=[(1-q)A_1z/(2A_2)]^{1/(q-1)}$. Changing to $z$ as the dependent variable, the above equation can be simplified to 
\begin{equation}
    z {d^2j_\pm\over d z^2}+(b_\pm-z){dj_\pm \over dz}-a_\pm j_\pm=0,
\end{equation}
where $b_\pm=2s_\pm/(q-1)$ and $a_\pm=(2+s_\pm)/(q-1)$. This is Kummer's differential equation \citep[e.g.,][]{Abram1972} which, provided $b$ is not a non-positive integer, has two independent 
solutions, which can be expressed in terms of the confluent hyper-geometric function as 
${_1F_1}(a_\pm,b_\pm;z)$ and $z^{1-b_\pm}{_1F_1}(1+a_\pm-b_\pm,2-b_\pm;z)$.

It is readily shown that the four solutions are not linearly independent since (for fixed $w,q$)
\begin{equation}\gamma^{s_\mp}{_1F_1}(a_\mp,b_\mp;z) \propto \gamma^{s_\pm}z^{1-b_\pm}{_1F_1}(1+a_\pm-b_\pm,2-b_\pm;z),
\end{equation}
and thus only 2 solutions need be retained, as is expected.

The general solution may be expressed simply as
\begin{equation}
n(\gamma)= C_+ \gamma^{s_+}{_1F_1}(a_+,b_+;z)+C_- \gamma^{s_-}{_1F_1}(a_-,b_-;z).
\end{equation}

For physically meaningful solutions, we require $n(\gamma\gg \gamma_{\rm max})\to0$. Using the asymptotic expression for large $z$ of the confluent hypergeometric function it follows that the solution decays as a power law $\propto \gamma^{-2}$ unless
\begin{equation}
 \frac{C_+\Gamma(b_+)}{\Gamma(b_+-a_+)}\gamma^{s_+}|z|^{a_+}+\frac{C_-\Gamma(b_-)}{\Gamma(b_--a_-)}\gamma^{s_-}|z|^{a_-}=0
\end{equation}
where $\Gamma(x)$ is the Gamma function. With this condition, the solution cuts off exponentially above $\gamma_{\rm max}$. This completes the derivation.

Finally, we consider the singular case of $q=1$, for which the solution is easily shown to be
\begin{equation}
n(\gamma)=C_1 \gamma^{\sqrt{w+(2-A_2/A_1)^2}-A_2/A_1}+C_2\gamma^{-\sqrt{w+(2-A_2/A_1)^2}-A_2/A_1}.\label{solq=1}
\end{equation}
Note that in this case the radiation-limited maximum Lorentz factor cannot be defined, e.g. see Eq.~(\ref{eq:gmaxsyn}). Although as discussed in the text, the solution is not meaningful when $r_{\rm L} > \Lambda_{\rm max}$.

%\listofchanges
\bsp	% typesetting comment
\label{lastpage}
\end{document}